\pdfoutput=1
\documentclass{JINST}
\usepackage{mathtools}
\usepackage{textcomp}
\usepackage{lineno}
\usepackage{multirow}
\usepackage{comment}

\title{Performance of new silica aerogels in a \\threshold \u{C}erenkov
counter}
\author{D.~Blyth$^a$\thanks{Corresponding Author},
R.~Alarcon$^a$, R.~Begag$^b$, J.~Holmes$^a$, and J.~Stryker$^a$\\
\llap{$^a$}Department of Physics, Arizona State University,\\ Tempe, AZ
85287-1504, U.S.A.\\ E-mail: \email{dblyth@asu.edu}\\\llap{$^b$}Aspen Aerogels,
Inc.,\\ Northborough,
MA 01532, U.S.A.}
\abstract{
New highly transparent, hydrophobic silica aerogels with refractive indices of
1.01 to 1.07 have been produced by Aspen Aerogels, Inc., and select tiles have
been tested using an electron beam at the DESY, Hamburg facility.  A
diffusively-reflective threshold \u{C}erenkov counter was designed and
constructed for the purpose of evaluating the aerogels, and can accommodate
aerogel tile areas of up to 20~cm by 20~cm.  Measurements of the performance of
the counter using the new aerogels as \u{C}erenkov radiators are given in terms
of photoelectron yields and a figure of merit.
}

\keywords{Particle identification methods, Cherenkov detectors}

\begin{document}
\section{Introduction}
Silica aerogels perform a vital role in nuclear and particle physics
experiments as radiators for \u{C}erenkov detectors.  These aerogels cover a
range of refractive index from about 1.01 to 1.08, and can be both highly
transparent and hydrophobic.  The often-large volume of aerogel required for
experiments can be very costly, and commercial alternatives to products such as
those produced by Matsushita Electric Works could be beneficial.  The work
presented in this paper is part of an effort to test the performance of new
hydrophobic aerogels produced by Aspen Aerogels in prototype \u{C}erenkov
detectors, and answers the call for proposals found in the U.S. DOE Small
Business Innovation Research (SBIR) Funding Opportunity Annoucements (FOAs)
(see for example page 103 of Phase I Release 1 found in \cite{SBIRFOAs}).  The
SBIR FOAs call for cost-effective \u{C}erenkov radiators with indices between
those of gases and liquids, e.g., aerogels.  The ultimate goal is the
qualification of commercially produced, cost-efficient, and high-quality
aerogel for use in the physics community by benchmarking the performance of the
aerogel in both threshold and ring-imaging \u{C}erenkov detectors.  In this
paper, we report on aerogel production and testing performed using a
threshold-type counter.

A number of aerogel tiles have been produced by Aspen Aerogels, and a small
selection have been used for testing in the research presented here.  A summary
of the techniques used in the production of the aerogels is given in section~\ref{SectionAerogel} along with a description of the technique used for
measuring the refractive index of the tiles.  In this section, we also present
the distribution of refractive index measurements taken.

The performance testing was achieved by measuring the performance of a
prototype \u{C}erenkov threshold counter using the new aerogels.
Quantification of the performance is given in terms of both photoelectron yield
from the photomultiplier tubes (PMTs) that view the counter, as well as
calculated figures of merit.  This approach is an alternative to measuring
individual optical properties of the aerogel, and instead aims to provide a
benchmark in the context of \u{C}erenkov particle detection.  The prototype
counter was evaluated using an electron test beam at DESY, Hamburg.  Details of
the experimental setup and test beam can be found in section~\ref{SectionSetup}.

Analysis of the experimental data is discussed in section~\ref{SectionAnalysis}, where a variation on a common and robust method of
photoelectron quantification is presented.  The analysis focuses on
determination of the relative inefficiency of detecting electrons in order to
extrapolate to average photoelectron yield using the application of Poisson
statistics.  This avoids difficulties often encountered when complex PMT
responses are fit.  Another desireable feature of this approach is that it
provides a means of measuring systematic error in the determination of
photoelectron yields.

In section~\ref{SectionFoM} we describe a figure of merit (FoM) intended to
quantify performance in a way that is comparable to different geometries.  This
FoM is derived from, and nearly identical to, that described by D.W.
Higinbotham \cite{Higinbotham1998332}, and as such provides a direct comparison
to previous experimental results analyzed in \cite{Higinbotham1998332}.  The
aerogel configurations tested, with tile sizes up to 20~cm by 20~cm, are
described in section~\ref{SectionResults} along with photoelectron yields and
figures of merit.  In addition, dependence of the photoelectron yields on
aerogel depth and incident beam position are presented and discussed.
\section{Silica aerogels}
\label{SectionAerogel}
\subsection{Sol-gel chemistry}
Fabrication of aerogels generally involves two major steps: formation of a wet
gel, and drying of the wet gel to form an aerogel. The vast majority of silica
aerogels prepared today utilize silicon alkoxide precursors; this route avoids
the formation of undesirable salt by-products and allows more control over the
final product. The most common alkoxide precursors are tetramethylorthosilicate
(TMOS) and tetraethylorthosilicate (TEOS), however, many others can be used to
impart different properties to the gel. The reaction is typically performed in
ethanol, with the aerogel density dependent on the concentration of alkoxide.
Hexamethyldisilazane (HMDZ) is often used during aging of TEOS or TMOS gels to
make the silica surface hydrophobic. However, this treatment takes time, and the
transparency of the aerogel is affected by the amount of HMDZ. In this program,
a methylsiliconate precursor was used as the co-precursor with TEOS, which
showed great results in terms of aerogel hydrophobicity and transparency.
\subsection{Aging process}
It is often assumed that the hydrolysis and condensation reactions of the
silicon alkoxides are complete when a sol reaches the gel point, however the
silica backbone of the gel still contains a significant number of unreacted
alkoxide groups. In fact, hydrolysis and condensation reactions continue after
several days beyond the gelation point. Failure to realize and accommodate this
is a common mistake preparing silica aerogel monoliths, as weak aerogels will
result.
Strengthening the gel is a result of the aging process, and is enhanced by
controlling the pH and water content of the aging bath. Common aging procedures
for base catalyzed gels involve soaking the gel in an alcohol/ammonia mixture,
at a pH of 8-9 for up to 48 hours at 60 \textdegree C. The time required for
this process depends on the gel thickness.  Any water left in the gel will not
be removed by supercritical drying, and will lead to an opaque, white, and very
dense aerogel.
\subsection{Supercritical drying}
The final process step for silica aerogels is supercritical drying. This is
where the liquid within the gel is removed, at supercritical conditions, leaving
only the linked silica network \cite{doi:10.1021/ja01250a034},
\cite{supercritfluidextract}. The process can be performed by venting the
ethanol above its critical point or by prior solvent exchange with CO$_2$
followed by supercritical CO2 venting \cite{tewari1986hydrolysis},
\cite{Tewari1985363}.  The gels are put in a high pressure vessel with liquid
CO$_2$ and solvent exchange occurs between ethanol and CO$_2$.  Once the ethanol
is extracted, the gels are brought to the critical point of CO$_2$ (31.06
\textdegree C and 1050 psi) and CO$_2$ is slowly vented.
Once the pressure reaches ambient, the vessel is opened and crack-free aerogel
monoliths are obtained. In this program, a 60 liter vessel was used to fabricate
the aerogel tiles for optical characterization. Batches of different refractive
index were fabricated for \u{C}erenkov detection performance evaluation at
Arizona State University.
\subsection{Refractive index}
\begin{figure}[h] \begin{center}
	\includegraphics[width=4in]{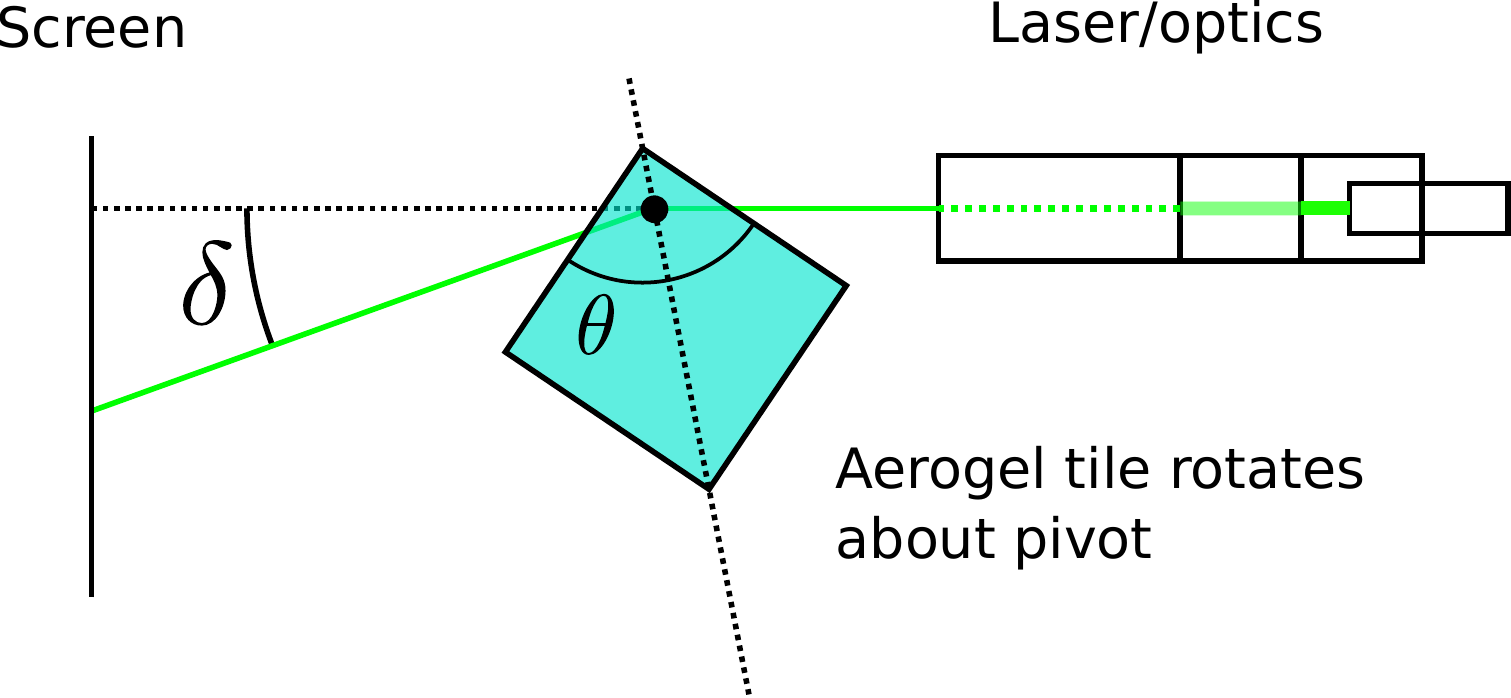}
\end{center}
\caption{A diagram representing the apparatus used to
	measure refractive indices of the aerogel tiles - The method of measuring the
	index of refraction for a tile of aerogel makes use of the simple prism effect. 
This method is very precise for low indices such as those of aerogel.}
\label{fig:IoRDiagram}
\end{figure}

Figure~\ref{fig:IoRDiagram} shows the setup for measuring the index of
refraction for a tile of aerogel.  The method is common (see
\cite{aThresholdCerenkov} and \cite{iormeasurement} for other accounts), and
easily yields precise measurements for low index materials like aerogel.
Refractive indices are calculated by measuring the minimum deflection of a laser
beam through the corner of an aerogel tile.  The relevant relationship is given
by the common formula for a prism,
\begin{equation}
	\label{eq:ior}
	n=\frac{\sin{\frac{\theta+\delta}{2}}}{\sin{\frac{\theta}{2}}},
\end{equation}
where $n$ is the refractive index at the wavelength of the laser,
$\delta$ is the minimum angle of deflection, and $\theta$ is the prism angle
(in this case, about $\pi/2$).  The minimum angle is found by measuring the
minimum deflection of the laser beam on a ruled screen.
The aerogel is pivoted about the intersection of the undeflected laser line, and
a line bisecting the corner of the tile.  As the aerogel is pivoted back and
forth, a minimum deflection is easily found.

The laser used is at a wavelength of 532 nm, and lies comfortably in the
spectral range of bialkali photocathodes, such as those used by the
photomultiplier tubes in our \u{C}erenkov test counter.

\begin{figure}[h] \begin{center}
	\includegraphics[width=4.5in]{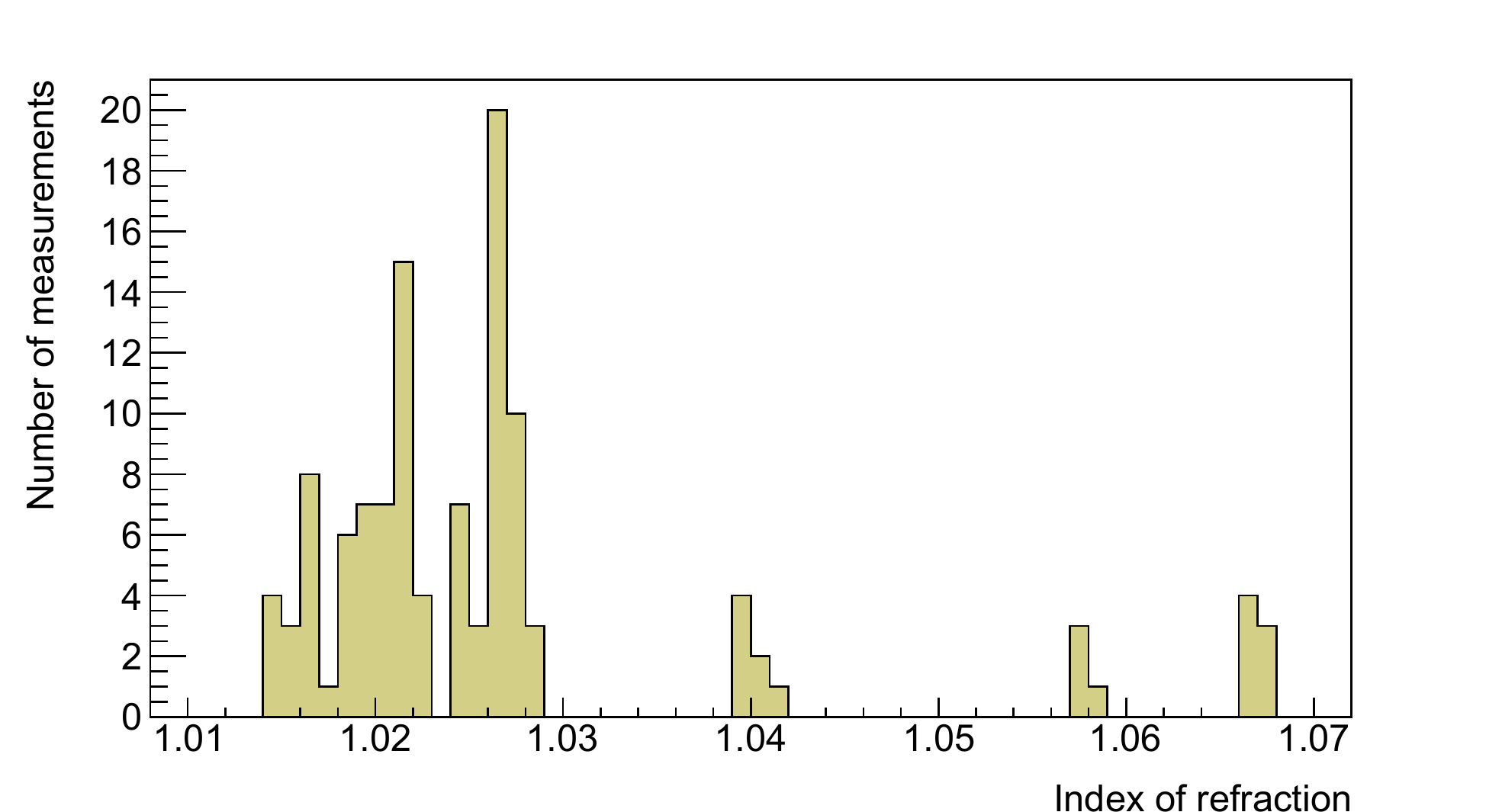}
\end{center}
\caption{Histogram of the refractive index
	measurements - For each tile, the refractive index was measured using the above
	method on more than one corner of the tile.  The histogrammed data represent
the range and distribution of indices produced for evaluation.}
\label{fig:IoRHisto}
\end{figure}

In Figure~\ref{fig:IoRHisto}, a histogram of all of the refractive index
measurements represents the range of indices produced for evaluation.  In this
paper, measurements of only a portion of the tiles are presented.
\section{Experimental setup}
\label{SectionSetup}
\subsection{\u{C}erenkov Counter}
\label{SectionCounter}
\begin{figure}[h]
	\begin{center}
		\includegraphics[width=4in]{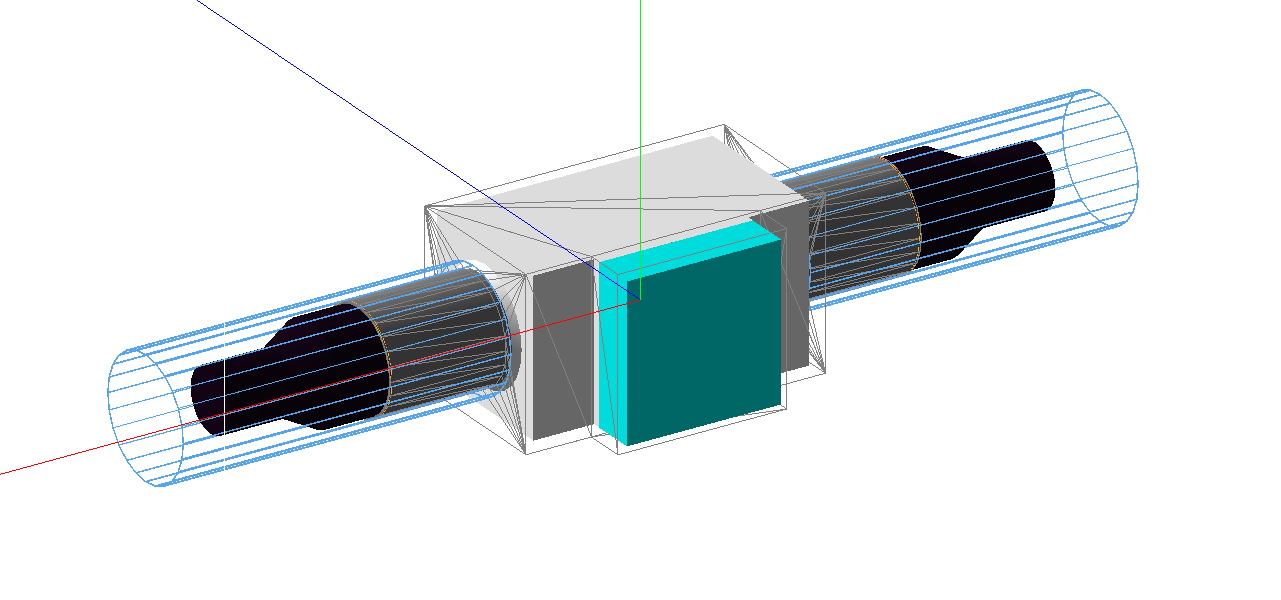}
	\end{center}
\caption{Counter Geometry - The solid filled regions in this image
	represent aerogel (cyan), air surrounded by diffuse reflector (white),
	aluminum light guides (grey), and PMT mock-ups (black).
	Represented in wireframe, the aluminum housing (grey) and magnetic
	shielding (light blue) are also shown.  The red, green,
	and blue axis lines
	represent the $\hat{x}$, $\hat{y}$, and $\hat{z}$ axes respectively, where the
	$\hat{z}$-axis is parallel to the beam.}
	\label{FigurePrettySimulation}
\end{figure}

A diffusively reflective threshold counter was developed for testing the new
aerogels.  It consists of an aluminum light box with dimensions 36.32 x 20.32 x
20 cm$^3$ in the $\hat{x}$, $\hat{y}$, and $\hat{z}$ directions respectively
(see Figure~\ref{FigurePrettySimulation} for axes), a 20.32 x 20.32 x 6.4 cm$^3$
box extruded upstream from the light box to house aerogel, and two 5'' Photonis
XP4500B photomultiplier tubes (PMTs) coupled to the sides of the light box.  A
0.6 mm-thick aluminum window is seen by the beam before passing through the
aerogel.  The light box and aerogel box are lined on the inside with a white,
diffusively reflective ``paper'', Millipore GVHP00010.
The PMTs are coupled to the light box with polished circular-cylindrical
aluminum light guides with an inside diameter of 13.6 cm.  The PMTs use a UV
glass and a bi-alkali photocathode, and manufacturer's specifications for the
Photonis XP4500B state a peak quantum efficiency of 24\%, with a spectral range
of about 200 - 650 nm where the efficiency falls off below 10\% of peak.

The purpose of the aluminum light guides is to couple the PMTs at a distance so
that shielding for static magnetic fields can be effectively used.  The test
counter is not tested in a significant magnetic field here, but the performance
of the counter itself in this configuration is of interest for future
applications.
In this setup, magnetic shields produced by MuShield are in place which reduce
ambient magnetic field at the photocathodes by a factor of 100.
\subsection{Data acquisition}
\label{SectionDAQ}
For readout, we use a 12-bit 250 MS/s digitizer (CAEN V1720).  The PMTs coupled
with the Photonis VD105k bases produce a signal with a FWHM $\approx4$ ns.  This
is too sharp of a pulse for the digitizer sample rate, so the PMT signals are
fed through passive 32 MHz low-pass LC filters.
When an external trigger is received, a number of digitized samples recorded in
the vicinity of the trigger are stored for offline analysis.  These samples
include data prior to the trigger, so that a pedestal can be calculated for each
event.
\subsection{Test beam}
\label{SectionTestBeam}
The T22 e-/e+ test beam at the DESY facility in Hamburg was used to perform
tests of the aerogel/counter combination.  The beam is generated by converting
bremsstrahlung radiation from carbon fiber targets in the DESY II lepton storage
ring.  The leptons in this storage ring are bunched, and revolve once every
microsecond \cite{desytestbeam}.
For this setup, converted electrons at an energy of 2 GeV are selected, with an
energy spread $\approx5\%$.
The beam is shaped by a 20 x 20 mm$^2$ primary collimator, and then a 6 x 6
mm$^2$ secondary collimator.  Located immediately downstream of the secondary
collimator is a broad scintillator (S1).  Approximately 5 m further downstream
is the \u{C}erenkov test counter, and another 5 m downstream is a 2 x 2 cm$^2$
scintillator (S2).  The test counter was mounted to an X-Y translation table
for relative beam positioning, and to measure position dependence.  A
trigger signal is produced with coincidence between S1 and S2.
The average trigger rate is $\approx4$ kHz.

For simplicity, it's beneficial to argue that on each trigger, only one electron
passes through the counter.  Given the trigger rate and the bunched nature of
the storage ring, we estimate that there are an average of 1.002 electrons
passing through the counter on triggered events.  This estimate relies on the
assumption that all electrons passing through the counter produce a trigger. 
Based on the observed narrowness of the beam, this is a reasonable assumption.
\section{Signal analysis}
\label{SectionAnalysis}
\subsection{Pulse binning}
\label{SectionPulseBinning}
\begin{figure}[h] \begin{center}
	\includegraphics[width=4in]{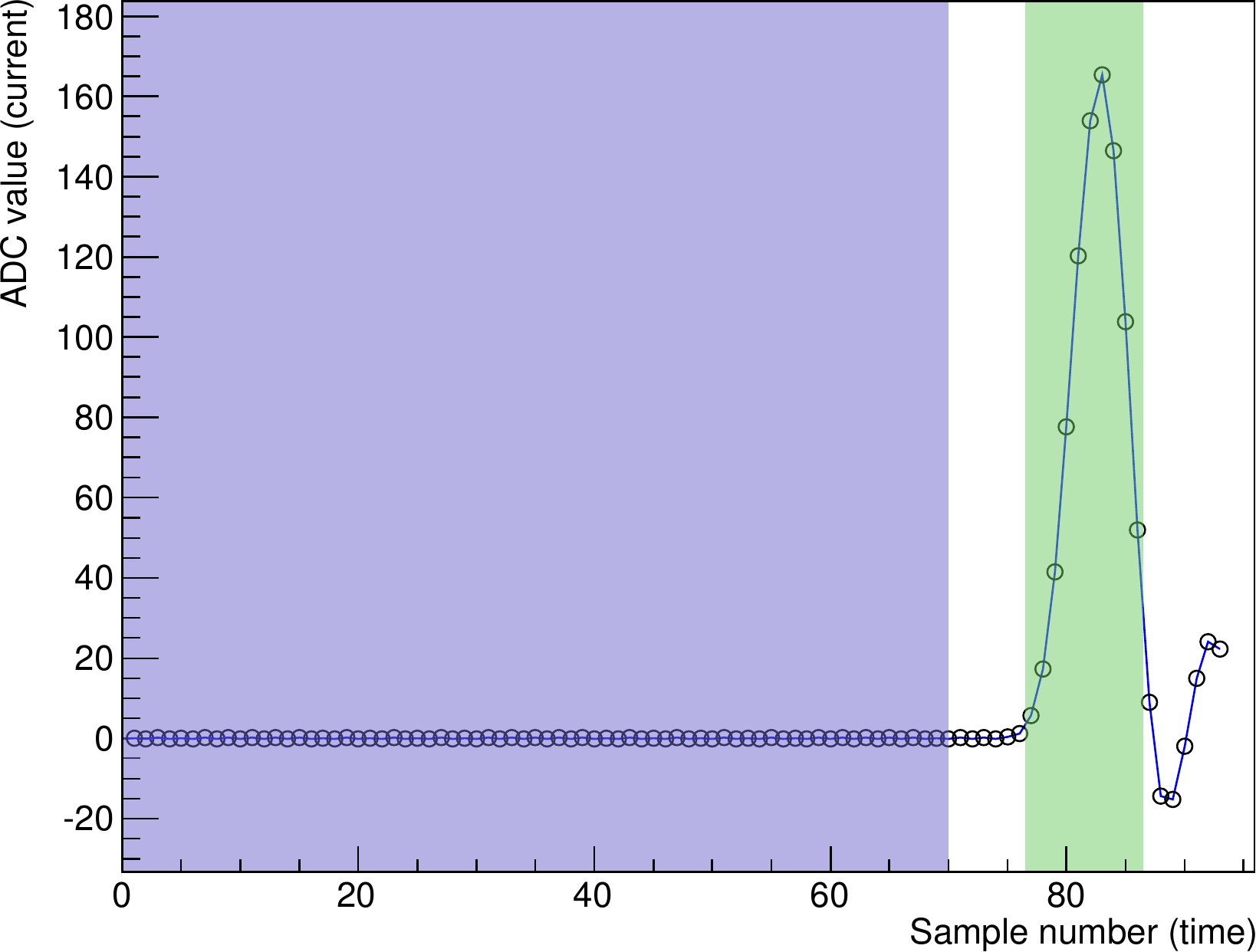}
\end{center}
\caption{PMT Response - An average digitized PMT
	pulse is shown with the height inverted, so that the pulses discussed here are
	positive.  The region in blue is used to calculate the electronic pedestal for
	each event.  This allows long-term variations in pedestal to be ignored.  The
	region in green is integrated relative to the pedestal for a value proportional
to the PMT anode charge.}
\label{FigureWaveform}
\end{figure}
\begin{figure}[h]
	\begin{center}
		\includegraphics[width=4in]{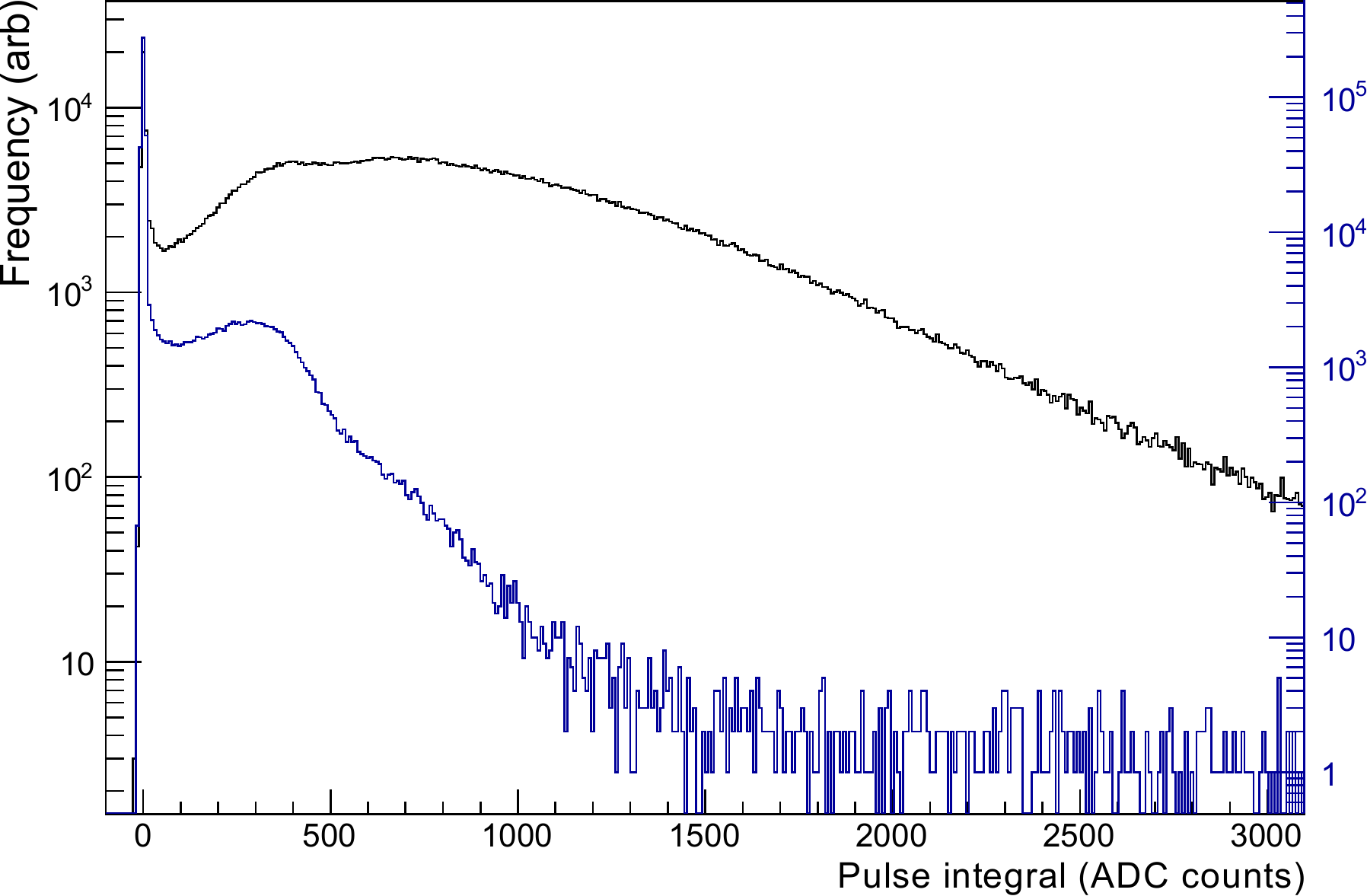}
	\end{center}
\caption{Pulse-Height Spectrum - The data shown are generated
	by binning the measured PMT charge for each triggerable event.  The
	black line represents a measurement with aerogel in the counter, while
	the blue line represents a measurement with only air in order to
	identify a single photoelectron peak.  The key features are the
	zero-photoelectron (pedestal) peak on the left, and the single- and
	multiple-photoelectron bumps to the right.  The pedestal peak falls off
	neatly to zero on the left in a Gaussian fashion, as can be seen in
	Figure~\protect\ref{FigureDeconvFit}.}
	\label{FigureSpectrum}
\end{figure}

As described in section~\ref{SectionDAQ}, the PMT waveforms are digitized to be
analyzed offline.  The timing of the signal is tuned so that 70 samples of
electronic pedestal are obtained before the desired PMT signal
(Figure~\ref{FigureWaveform}).  The pedestal for each event and for each PMT is
therefore amply measured in order to produce a sharp zero-photoelectron peak,
on which our analysis ultimately relies.  For each event, the latter portion of
the waveform in Figure~\ref{FigureWaveform} is integrated relative to the
calculated pedestal and binned into a pulse-height spectrum
(Figure~\ref{FigureSpectrum}).
\begin{figure}[h]
	\begin{center}
	\includegraphics[width=4in]{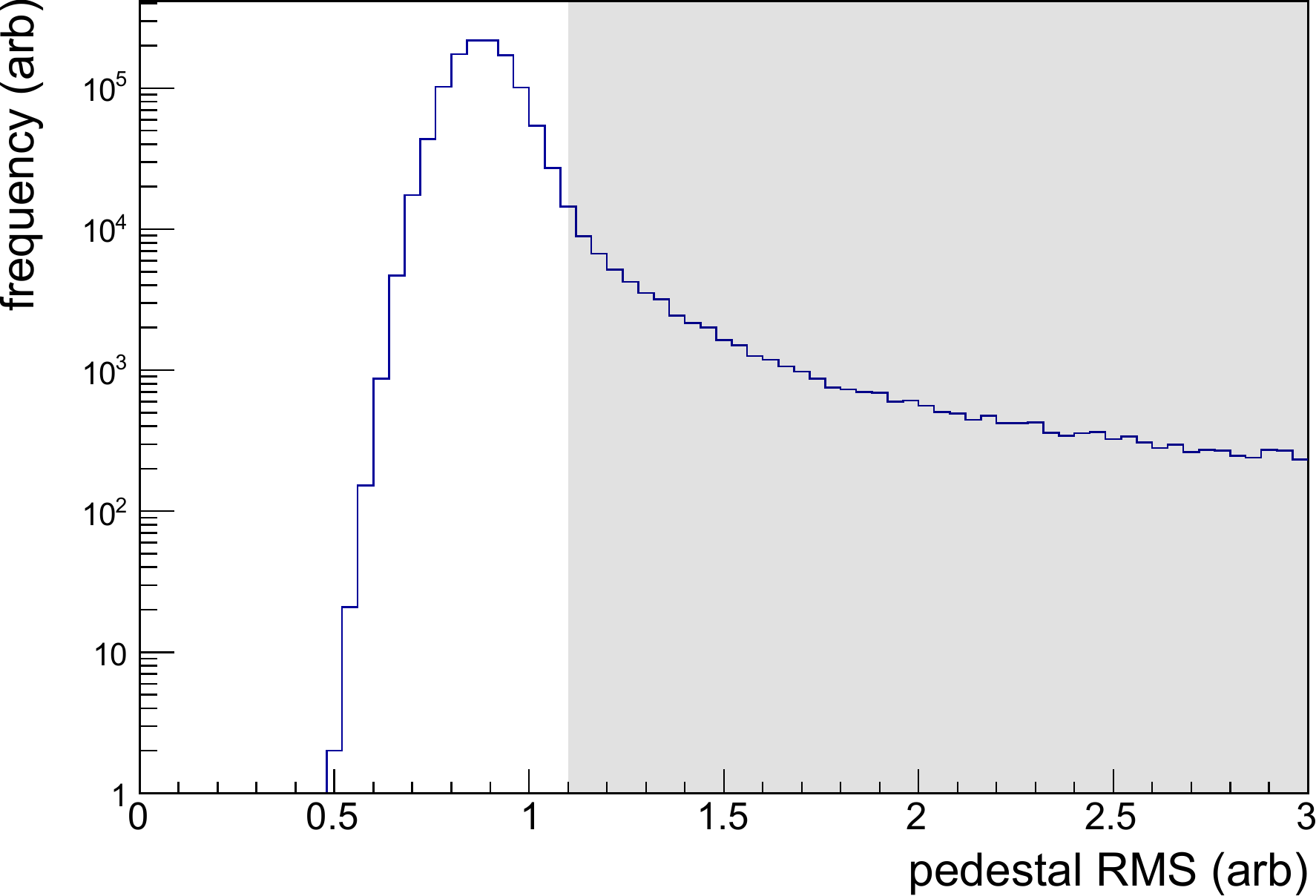}
\end{center}
\caption{Histogram of pedestal RMS values - The desirable
	Gaussian-like peak of the distribution continues into a region where the tails
	of thermal dark current pulses begin to dominate the RMS.  This produces artifacts in the
	pedestal calculation which assumes high-frequency stochastic fluctuations. 
	Therefore, a cut is placed on the pedestal RMS (the rejected events shown by the
	grey region) where the distribution fails to be dominated by such fluctuations.
	The calculated photoelectron yield has negligible sensitivity to the placement
of this cut in the range of 1.0 - 1.2.}
\label{FigurePedCut}
\end{figure}
In order to minimize the effects of dark current, a cut is placed on the
standard deviation of the pedestal samples to enforce the assumption of a
constant signal with a small random error.  This cut, shown in
Figure~\ref{FigurePedCut}, is necessary for minimizing artifacts in the
zero-photoelectron peak due to dark current.
\subsection{Determination of average photoelectron counts}
The pulse-height spectrum shown in Figure~\ref{FigureSpectrum} has a sharp peak
at the left corresponding to the absence of photomultiplication, and its
integrated value divided by the total number of events is therefore considered
to be the probability that zero photoelectrons are produced in the PMT for a
triggerable event in a given counter configuration.  The probability
distribution for the number of photoelectrons emitted from the PMT cathodes in a
triggerable event is well approximated by a Poisson, given the low
probability of a radiated photon producing a photoelectron.  We therefore claim
that the average number of photoelectrons produced for such an event is equal to
\begin{equation}
	\label{EqPoissonSolution}
	\left<N_{PE}\right> = -\ln{\frac{N_{ped}}{N_{tot}}}
\end{equation}
where $N_{ped}$ and $N_{tot}$ are the pedestal event and total event counts
respectively.  This general approach has been compared to other techniques in
\cite{dossi_methods_2000}, where R.~Dossi et al. also incorporate a model for
single- and multiple-photoelectron peaks for a $\chi^2$ fit to data.  We have
chosen this approach since it has the advantage of few assumptions, and we have
found it to be robust in our application.

Systematic error in the determination of $N_{ped}$ comes from overlap
of the pedestal peak with contributions from non-zero photoelectron signals.  This
is potentially largely due to incomplete photomultiplication in the large PMTs.
In practice, a sharp pedestal peak is necessary and a cutoff is placed on
the integration of the pedestal peak and varied to examine systematic effects. 
As a guide for this cutoff, a method of determining the width of the pedestal
peak will be incorporated in order to place the cutoff at an estimated
probability of pedestal noise excursions.
\subsubsection{Deconvolution}
\begin{figure}[h]
	\begin{center}
		\includegraphics[width=4in]{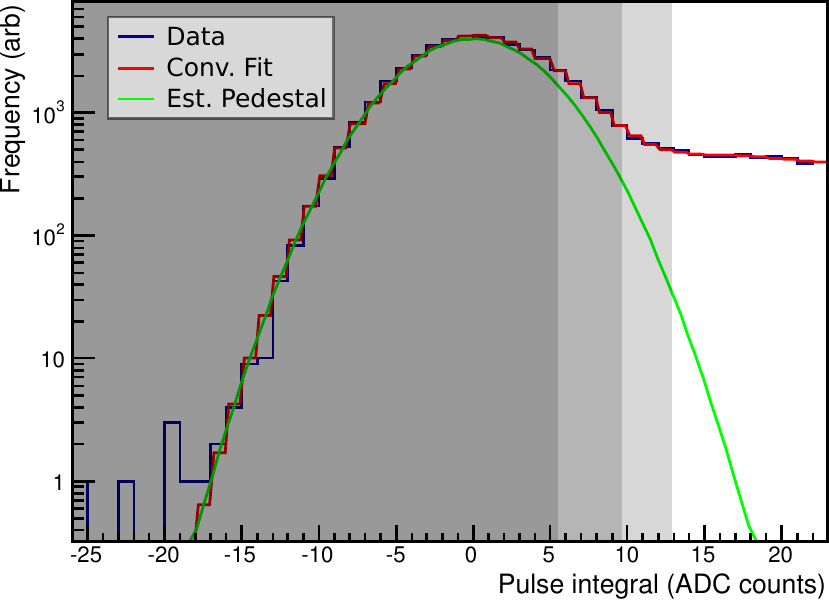}
	\end{center}
\caption{Convolution Fit - The data shown are from the
	zero-photoelectron (pedestal) peak on the left side of Figure~\protect\ref{FigureSpectrum}.
	The red line represents a reconvolution of the deconvolved spectrum
	which has been fit to the data to extract the width of the PSF (point-spread
	function).
	The reduced $\chi^2$ of the fit is $102/97$.  The green line
	shows the zero pulse-height contribution from the deconvolved spectrum.
	The three grey regions illustrate the pedestal integration to 90\%,
	99\%, and 99.9\% of the total probability (see
	Equation~\protect\ref{EqPedIntegral}).}
	\label{FigureDeconvFit}
\end{figure}
To determine the width of the pedestal peak, a deconvolution method is applied to the
pulse-height spectrum with an estimated point-spread function (PSF). The PSF is
approximated by a Gaussian with an unknown width, and the reconvolved spectrum
is fit to the measured spectrum using a least-squares search method over the PSF
width parameter.
Figure~\ref{FigureDeconvFit} gives an example of the result of such a fit.  The
purpose of such a process is to gain a precise form of the PSF.

By constraining the deconvolved spectrum to exist in a positive semi-definite
range, we can demand that the true charges deposited on the PMT anode are of only one
sign (negative, since the sign is inverted for a positive spectrum). This allows
the pedestal region of the spectrum to provide a stable constraint on the PSF.
In this way, the fitted PSF can be said to represent the random error in
the measurement of charge collected by the PMT anode.

An iterative deconvolution approach known as the Lucy-Richardson algorithm is
used \cite{citeulike:1719872}.  This approach provides a straightforward means
of obtaining a maximum-likelihood solution while enforcing the condition $\psi_i
\geq 0$, where $\psi_i$ represent the discrete values of the deconvolved
spectrum.
\subsubsection{Pedestal integration}
Figure~\ref{FigureDeconvFit} shows the pedestal peak of a pulse-height spectrum
along with the predicted pedestal contribution from the deconvolved spectrum.
To determine the pedestal contribution from the data, the pulse-height spectrum
is integrated from left to right to specific probability intervals based on the
fitted PSF.  The pedestal integral is taken to be
\begin{equation}
	\label{EqPedIntegral}
	N_{ped,i} = \displaystyle \int_{-\infty}^{\text{erfc}^{-1}\left[2(1-P_i)\right]\sqrt{2}
\sigma}
{\tilde\phi\left(x\right)dx}
\end{equation}
where $\text{erfc}^{-1}$ is the inverse complementary error function -- $P_i$ are 0.9,
0.99, and 0.999 -- $\sigma$ is the width of the Gaussian PSF --
and $\tilde\phi\left(x\right)$ is the measured spectrum.  By combining Equation~\ref{EqPoissonSolution} and~\ref{EqPedIntegral}, $P_i$ provide average
photoelectron values ($\left<N_{PE,i}\right>$) for a mid-range result at
99\%, and a reasonable range from the upper bound at 90\% and the lower bound
at 99.9\%.
\section{Figure of Merit}
\label{SectionFoM}
D.W. Higinbotham \cite{Higinbotham1998332} presents a useful Figure of Merit
for threshold
\u{C}erenkov counters in the form of
\begin{equation}
	\label{EquationHiginFoM}
	H_{D.W.H.} \equiv \frac{\left<N_{PE}\right>}
	{L\left(1-\frac{1}{\beta^2n^2}\right)}*\frac{1-\eta\left(1-\epsilon\right)}{\epsilon},
\end{equation}
where $L$ is the depth of the aerogel, $\eta$ is the average light box reflectivity
(using
reasonable values for reflector and aerogel reflectivity given by
\cite{Higinbotham1998332} for uniformity), and $\epsilon$ is the fraction of the
light box covered by photosensitive detector.
The factor $L\left(1-\frac{1}{\beta^2n^2}\right)$ is
approximately proportional to the number of photons radiated from the aerogel, and ${\epsilon}/\left[{1-\eta
\left(1-\epsilon\right)}\right]$ is the limit of
\begin{equation}
	\displaystyle
	\sum^{\infty}_{n = 0}{\epsilon\left[\eta\left(1-\epsilon\right)\right]},
\end{equation}
which estimates the probability of a photon
reaching a photocathode under the assumption that there is a flat distribution
of reflections across the surface of the light box.
$H$ then contains information on light losses in the aerogel, quantum efficiency of
photon detectors, and higher-order geometrical effects.
In the results of this paper, we will utilize this FoM in a slightly modified form.

Due to the depth of our counter and the rapidity of the beam used for testing,
\u{C}erenkov radiation in the air contributes non-trivially to our measured
photoelectron yield, making a FoM of the form of Equation~\ref{EquationHiginFoM}
artificially large. We will account for this by taking
\begin{equation}
	L\left(1-\frac{1}{\beta^2n^2}\right) \rightarrow L_g\left(1-\frac{1}{\beta^2n_g^2} \right) +
	L_a\left(1-\frac{1}{\beta^2n_a^2}\right)
\end{equation}
so that
\begin{equation}
	\label{EquationFoM}
	H \equiv \frac{\left<N_{PE}\right>}{L_g\left(1-\frac{1}{\beta^2n_g^2}\right)
	+ L_a\left(1-\frac{1}{\beta^2n_a^2}\right)}*\frac{1-\eta\left(1-\epsilon\right)}
	{\epsilon},
\end{equation}
where subscripts $g$ and $a$ represent aerogel and air respectively.  This
provides an $O\left(10\%\right)$ correction that reduces the FoM to compare more
directly with counters that have less contribution from air.
\section{Aerogel configurations and results}
\label{SectionResults}
\begin{table*}[h]
\caption{Aerogel configuration results - The table below lists for each
	aerogel configuration the batch number, total aerogel depth, average photoelectron yield,
	mean refractive index, standard deviation of index measurements, and a Figure of Merit
	(Equation~\protect\ref{EquationFoM})
	in the style of D.W. Higinbotham \cite{Higinbotham1998332}.}
	\label{TableResults}
	\begin{center}
		\begin{tabular}{c|c|c|c|c|c}
			Batch & Depth (mm) & $\left<N_{PE}\right>$ & $\left<n\right>$ & $\sigma_n$ & $H$
			(cm$^{-1}$) \\
			\hline
			Empty (air) & 0 & $0.58\substack{+0.19 \\ -0.02}$ & $\approx1.00028$ & N/A &
			N/A\\
			\hline
			\multirow{3}{*}{1} & 28 & $3.81\substack{+0.29 \\ -0.08}$ & \multirow{3}{*}{$1.026$}
			& \multirow{3}{*}{$0.0016$} & 40\\
			& 41 & $4.91\substack{+0.32 \\ -0.10}$ & & & 36\\
			& 56 & $6.04\substack{+0.34 \\ -0.12}$ & & & 32\\
			\hline
			2 & 53 & $6.36\substack{+0.34 \\ -0.12}$ & $1.028$ & 0.0007 & 34\\
			\hline
			3 & 62 & $5.60\substack{+0.28 \\ -0.09}$ & $1.028$ & 0.0023 & 28
		\end{tabular}
	\end{center}
\end{table*}
The performance of the \u{C}erenkov threshold test counter has been analyzed with three
different batches of aerogel.  The batches provide a
small spread in refractive index and two different aerogel tile form factors.
The tiles from each batch are 1 - 2 cm deep squares.
\begin{description}
	\item[Batch 1] has a mean refractive index of
		1.026 and has a square area $\approx\text{10 x 10 }\text{cm}^2$.
	\item[Batch 2] has a mean refractive index of 1.028, and also has a square area
		$\approx\text{10 x 10 }\text{cm}^2$.
	\item[Batch 3] also has a mean refractive index of 1.028, but with a square area
		$\approx\text{20 x 20 }\text{cm}^2$.  This is considered a large tile.
\end{description}

\begin{figure}[h] \begin{center}
	\includegraphics[width=4in]{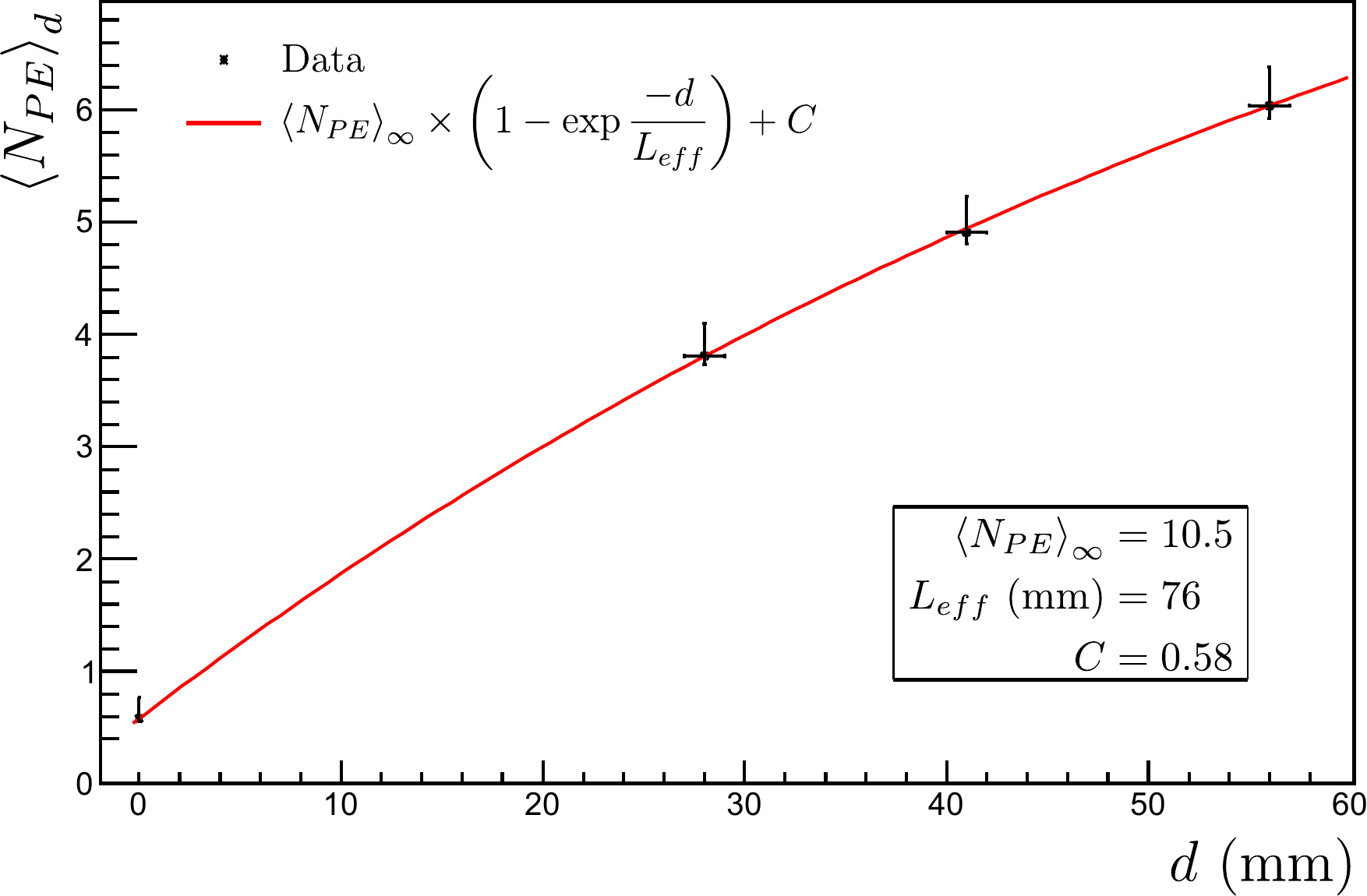}
\end{center}
\caption{Batch 1 depth dependence - The red curve
	below represents a fit to the data meant to extract the effective aerogel
	absorption length $L_{eff}$ and a maximum photoelectron yield for an arbitrarily
	deep stack of aerogel.  The $\chi^2$ value for the fit is 0.01, which is not
	unreasonable for a single degree of freedom, but tends to suggest that the
	systematic errors stated in Table~\protect\ref{TableResults} are at high confidence
intervals as intended.}
\label{FigureDepthDep}
\end{figure}

\begin{figure*}
	\begin{center}
	\includegraphics[width=4.5in]{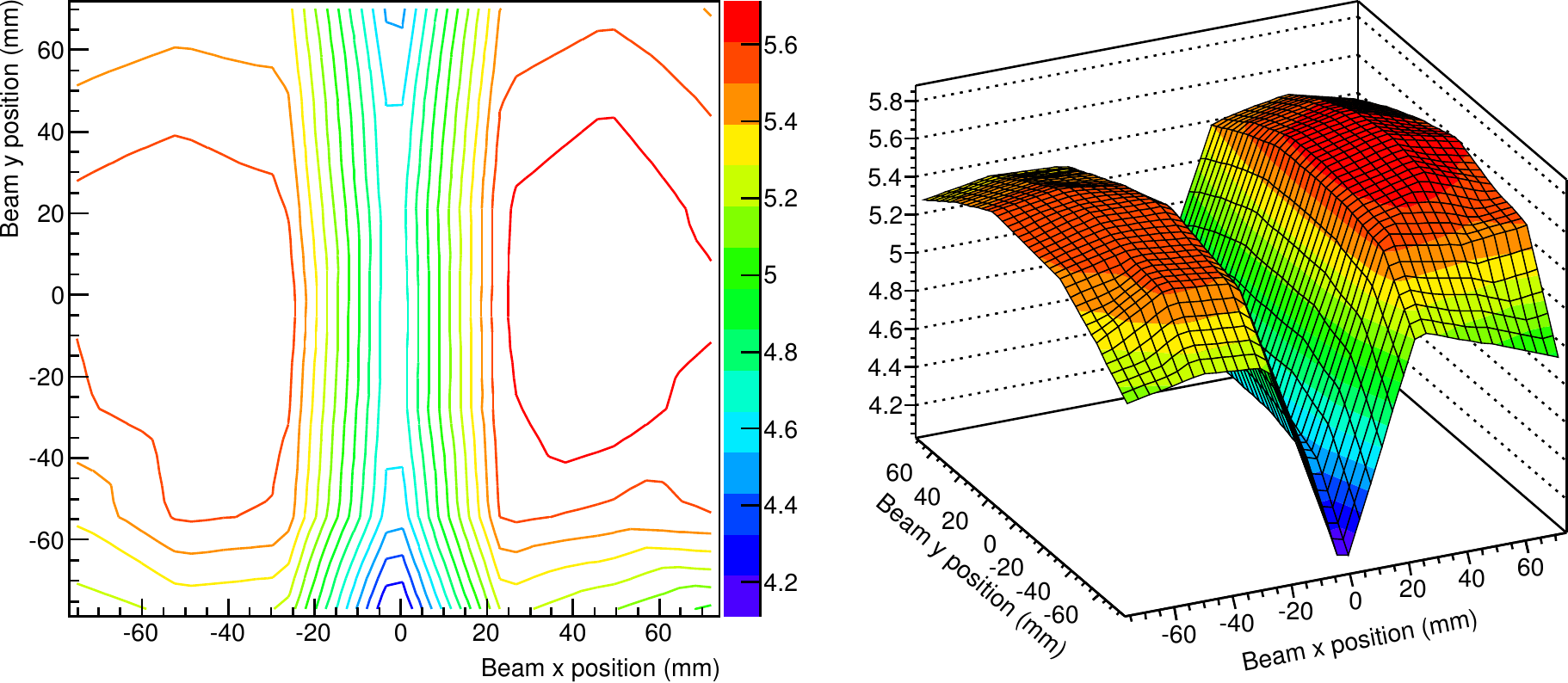}
\end{center}
\caption{Spatial scan - The spatial dependence of the average photoelectron
	yield
	$\left<N_{PE}\right>$ of the
	test counter with the Batch 3 large tile configuration is shown below.  A large dip
	down the
	center is clearly illustrated, which is due to a metal support in the test counter
	blocking
	the \u{C}erenkov light cone.
	A slight preference in the +x
	direction exists due to differences in quantum efficiency between the PMTs on
either side of the aerogel.}
\label{FigureSpatialScan}
\end{figure*}

Table~\ref{TableResults} lists measurements for a series of aerogel configurations.
For each configuration, the total aerogel depth, average number of photoelectrons
$\left<N_{PE}\right>$, the mean refractive index, the spread in index measurements,
and a Figure of Merit have been calculated.
Figure~\ref{FigureDepthDep} shows a plot of the first four configurations listed in
Table~\ref{TableResults}, along with a fit to the values showing the non-linear depth
dependence of Batch 1.
This fit is of the form used by I. Adachi et al.
\cite{Adachi1995390}, and is given by
\begin{equation}
	\label{eqFit}
	\left<N_{PE}\right>_d=\left<N_{PE}\right>_\infty\times\left(1-\exp{\frac{-d}{L_{eff}}}\right)+C,
\end{equation}
where $d$ is the depth of the aerogel stack, and the effective absorption
length $L_{eff}$ contains information about absorption and scattering
in the aerogel convolved with the geometry and reflectivity of the test counter.
The spatial dependence of the test counter response with the Batch 3 large tiles
has been measured by scanning the counter transversely with respect to the beam,
and is shown in Figure~\ref{FigureSpatialScan}.
\subsection{Large tile performance}
Examining the features in Figure~\ref{FigureSpatialScan}, we find that a 1
cm-wide strip of aluminum that was used to hold the aerogel tiles in place
during testing produced a significant effect in the spatial scan as indicated by
the large dip in the center of the scan.  There is also a small asymmetry in the
response of the counter on either side of the aluminum support, indicating a
difference in quantum efficiencies between the two PMT photocathodes.  Finally,
a gentle slope exists toward the edges of the scan as a result of the geometry
of the light box.  A Monte-Carlo simulation was performed which qualitatively
replicated each of these features.

The spatial scan shows that the uniformity of the counter in the regions not
affected by the aluminum aerogel support is quite good.  Each of the
measurements listed in Table~\ref{TableResults} were obtained with the beam
center varying by no more than a few mm relative to the counter, and are each
around 50,-50 mm.  We conclude that the small variations in relative beam center
produced variations between batches in the average number of measured
photoelectrons on the order of a percent.
\section{Summary}
\label{SectionSummary}
Hydrophobic aerogels have been produced for use as \u{C}erenkov radiators in a
threshold counter.  Measurements of the performance of the aerogel/counter
combination with aerogel refractive indices from 1.026 to 1.028 and various
depths have been performed, yielding photoelectron counts between 3.8 and 6.4
(Table~\ref{TableResults}).  Using a figure of merit in the style of D.W.
Higinbotham \cite{Higinbotham1998332}, low-order effects from the specific
refractive indices used -- as well as from the geometry (aerogel depth, PMT
area, etc) -- have been removed to produce values from 28 - 40 cm$^{-1}$ that
can be used to compare to previous experimental results.
\acknowledgments
This material is based upon work supported by the Department of Energy under
Award Number DE-SC0004290 and by the National Science Foundation Awards 0969654
and 1306547.  The authors would also like to thank the OLYMPUS collaboration for
the beam time at the DESY detector testing facility.
\paragraph{Disclaimer}
This report was prepared as an account of work sponsored by an agency of the
United States Government.  Neither the United States Government nor any agency
thereof, nor any of their employees, makes any warranty, express or implied, or
assumes any legal liability or responsibility for the accuracy, completeness, or
usefulness of any information, apparatus, product, or process disclosed, or
represents that its use would not infringe privately owned rights.  Reference
herein to any specific commercial product, process, or service by trade name,
trademark, manufacturer, or otherwise does not necessarily constitute or imply
its endorsement, recommendation, or favoring by the United States Government or
any agency thereof.  The views and opinions of authors expressed herein do not
necessarily state or reflect those of the United States Government or any agency
thereof.
\bibliographystyle{JHEP}
\bibliography{my}
\end{document}